\documentclass[aps,pra,reprint,twocolumn,superscriptaddress,showpacs,showkeys,floatfix,bibnotes]{revtex4-2}
\usepackage{polski}
\usepackage{amsmath,amssymb,amstext}
\usepackage{ulem}
\usepackage[usenames,dvipsnames]{color}
\usepackage{bm}
\usepackage{mathtools}
\usepackage[x11names]{xcolor}
\newtagform{blue}{\color{RoyalBlue3}(}{)}
\usepackage{graphicx}
\usepackage{braket}
\usepackage{natbib}
\usepackage{comment}
\usepackage{dcolumn}
\usepackage[english]{babel}
\usepackage{wasysym}
\usepackage[colorlinks,bookmarks=false,citecolor=blue,linkcolor=red,urlcolor=blue]{hyperref}

\begin{document}
	
	\title{Nonlinear Thouless Pumping of Solitons Across an Impurity}
	
	\author{Xuzhen Cao}
	\affiliation{State Key Laboratory of Quantum Optics and Quantum Optics Devices, Institute of Laser Spectroscopy, Shanxi University, Taiyuan 030006, China}
	\affiliation{Collaborative Innovation Center of Extreme Optics, Shanxi University, Taiyuan, Shanxi 030006, China}

	\author{Chunyu Jia}
	\affiliation{Department of Physics, Zhejiang Normal University, Jinhua 321004, China}
	
	\author{Ying Hu}
	\thanks{huying@sxu.edu.cn}
	\affiliation{State Key Laboratory of Quantum Optics and Quantum Optics Devices, Institute of Laser Spectroscopy, Shanxi University, Taiyuan 030006, China}
	\affiliation{Collaborative Innovation Center of Extreme Optics, Shanxi University, Taiyuan, Shanxi 030006, China}

	\author{Zhaoxin Liang}
	\thanks{zhxliang@zjnu.edu.cn}
	\affiliation{Department of Physics, Zhejiang Normal University, Jinhua 321004, China}

	\begin{abstract}
	The nonlinear Thouless pumping is an exciting frontier of topological physics. While recent works have revealed the quantized motion of solitons in Thouless pumps, the interplay between the topology, nonlinearity and disorder remains largely unexplored. Here, we investigate the nonlinear Thouless pumping of solitons in the presence of an impurity in the context of a Bose–Einstein condensate. Using both the Gross–Pitaevskii equation and Lagrangian variational approach, we analyze the interaction between a moving soliton and an impurity. Without the pump, the soliton can pass through a light impurity, but gets trapped by the heavy impurity. In contrast, we find Thouless pump soliton in Thouless pumps can transit through also a heavy impurity, and its motion is topologically quantized. Our result explicitly showcases the robustness of topological soliton pumping against microscopic imperfections, and opens a new perspective in the  information processing with solitons.
	
	\end{abstract}
	
	\maketitle
	
	\section{Introduction} 
	
Central to understanding quantized topological transport is the concept of Thouless pumping~\cite{Thouless1982,Thouless1983,Niu1984,Di2010}. There, a quantum particle acted on by a periodic potential that varies adiabatically and periodically in time shows a quantized motion; the quantized displacement, dictated by the Chern number of the underlying band structure in the momentum-time space, is stable against disorder~\cite{Di2010}. The experimental implementation of Thouless pumps and the observations of topologically quantized transport have been achieved in a wide variety of systems, including ultracold atoms~\cite{Wang2013,Lohse2016,Nakajima2016,Lu2016,Lohse2018,Nakajima2021,Minguzzi2022}, photonics~\cite{Kraus2012,Yongguan2016,Zilberberg2018,Cerjan2020,WangPeng2022}, and spin systems~\cite{Wenchao2018,Bastidas2022}. Interestingly, these synthetic topological systems can operate even beyond the linear regime, i.e., in the presence of nonlinearities that arise from the Kerr effect in optical platforms~\cite{Leykam2016,Tangpanitanon2016,Bisianov2019,Ivanov2020,Tianxiang2021} or the interparticle interactions in atomic setups~\cite{Solnyshkov2017,Cuadra2020,Lo2021}. At present, exploring nonlinear topological pumping has attracted significant interests and experimental efforts. 
	
A paradigm of the nonlinear Thouless pumping concerns the topological transport of nonlinear excitations, known as solitons~\cite{Jurgensen2021,Mostaan2022,Qidong2022,Fu2022,Jurgensen2022,Jurgensen2023,Tuloup2023}. It has been shown that for weak nonlinearity, the motion of the soliton can be topologically quantized, where the quantized displacement is directly related to the topology of the underlying band structures~\cite{Citro2023,Kraus2012,Lohse2016,Nakajima2016,Yongguan2016,Lu2016,Wenchao2018,Cerjan2020,Nakajima2021,Minguzzi2022,Xiang2023}; whereas, the quantization breaks down in the strong nonlinearity limit~\cite{Niu1984,Berg2011,Qian2011,Wang2013,Grusdt2014,Zeng2016,Tangpanitanon2016,Li2017,Ke2017,Taddia2017,Hayward2018,Nakagawa2018,Stenzel2019,Haug2019,Unanyan2020,Greschner2020,Chen2020,Qidong2022,Mostaan2022}. More recent studies have shown that the nonlinearity can fundamentally modify the adiabatic dynamics and thus the Berry connection~\cite{Tuloup2023}. So far, however, exploring the interplay of nonlinear excitations, topology and disorder remains largely unchartered territory. 

In this work, we explore the soliton transport in nonlinear Thouless pumps in the presence of an impurity, based on a quasi-one-dimensional (1D) Bose-Einstein condensate (BEC). We first numerically simulate the motion of the soliton through the impurity using the Gross–Pitaevskii (GP) equation~\cite{Perez1996,Dalfovo1999,Khawaja2002}. Further, we analytically study the interaction between the soliton and the impurity using the Lagrangian variational approach. Without the pumps, the fate of the soliton relies crucially on the effective mass of the impurity: For light impurities, the soliton can easily pass through, but for impurities with large masses, the soliton can get trapped. However, in the presence of Thouless pumps, we find the soliton can transit through the impurity, and its motion becomes topologically quantized. Such topology-enhanced soliton transport through the impurity provides insights into the interplay between nonlinearity, topology and disorder, and explicitly elucidates the stability of the topological nonlinear pumping against local imperfections. Our result may enable novel possibilities in the information processing with solitons. 

The paper is organized as follows. In Sec.~\ref{MS}, we present detailed descriptions of our model system. In Sec.~\ref{NumericalGP}, by numerically solving GP equation, we investigate the nonlinear Thouless pumping of solitons in the presence of an impurity. Sec.~\ref{VariationalGP} we analytically derive the equations of motions for the soliton using Lagrangian variational approach, to gain more understandings of its interaction with the soliton. Both the numerical and analytical results explicitly demonstrate the robustness of topological soliton pumping against imperfections. In Sec.~\ref{Con} we summarize our work.

\section{Model system} \label{MS}

We consider a 3D BEC with an attractive interatomic interaction in the following trap geometry~\cite{Wang2013,Lohse2016}: in the $x$- direction, the BEC is trapped in an optical superlattice $V_{\text {OSL}}(x,t)$ and an impurity potential $V_{{\text {imp}}}$, while in the 
$y$- and $z$- directions, the BEC is tightly confined in the harmonic trap with the large trap frequency $\omega_\perp$ such that the atomic motions in these directions are affectively frozen.  At the mean-field level, the considered BEC can be described by the condensate wavefunction $\Psi(x,y,z)=\psi(x)\phi(y,z)$ with $\phi(y,z)=1/(\sqrt{\pi}a_{\perp})e^{-[y^2+z^2]/(2a_\perp^2)}$ and $a_\perp=\sqrt{\hbar/(m\omega_\perp)}$.
Averaging out the transverse degrees of freedom results in an effective quasi-1D BEC described by the following GP equation~\cite{Forinash1994,Jia2022}, i.e., 
\begin{equation}
\!\!\!	i\hbar\frac{\partial\psi}{\partial t}=-\frac{\hbar^2}{2m}\frac{\partial^2\psi}{\partial x^2}+g|\psi|^2\psi+[V_{{\text {imp}}}(x)+V_{{\text {OSL}}}(x,t)]\psi.\label{Eq:GPE1}
\end{equation}
Here the $m$ is the atomic mass, and $g=2 \hbar^2a_s/(ma_\perp^2)$ is the confinement-modified coupling constant, where $a_s<0$ is the negative s-wave scattering length. 

 \begin{figure}[t]
     	\centering
     	\includegraphics[width=1\columnwidth]{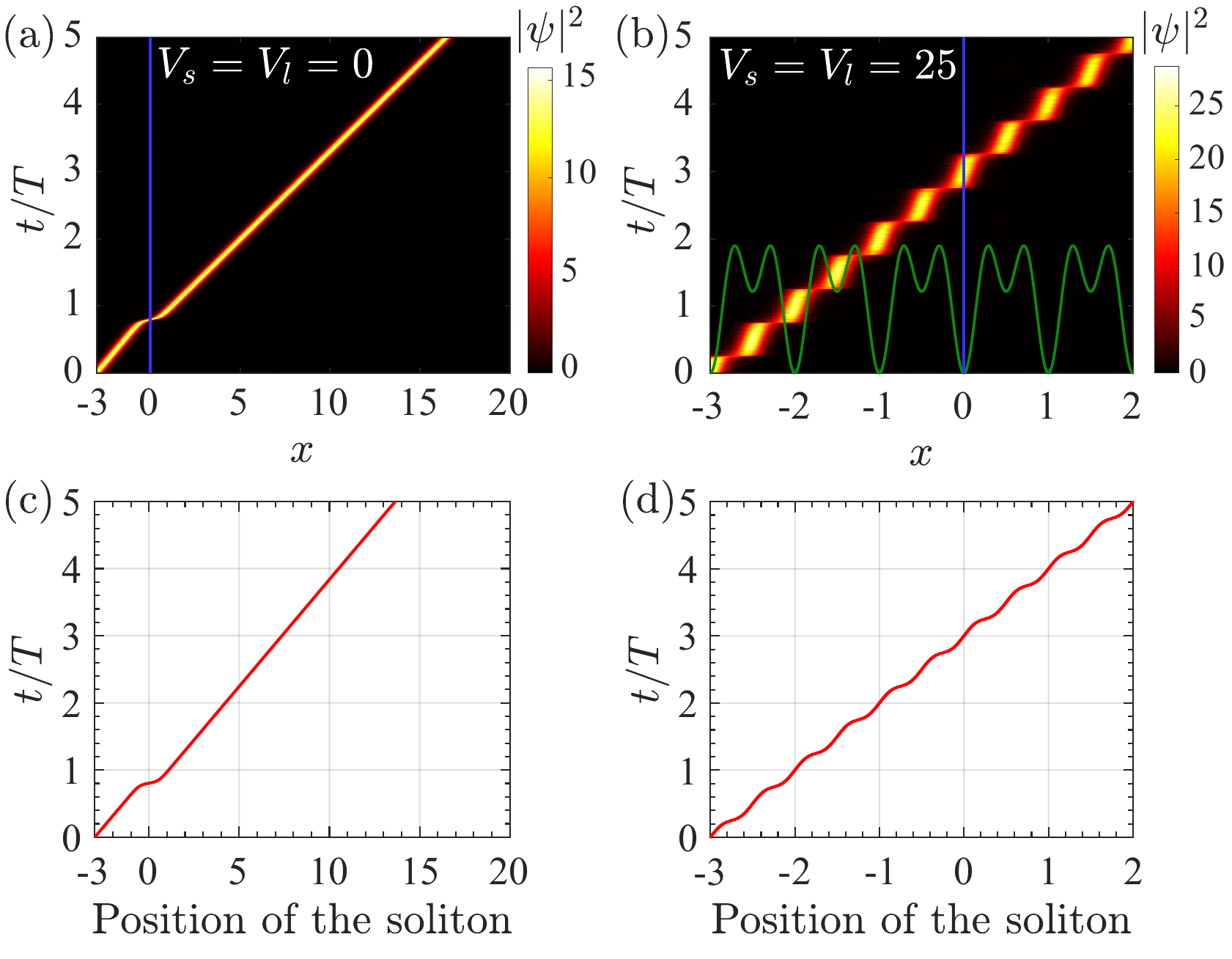}
     	\caption{Soliton transport in the presence of a light impurity without and with nonlinear Thouless pumps. In (a) and (c), the pump is absent, where $V_s=V_l=0$. In (b) and (d), the pump is present, with $V_s=V_l=25$ and the periodicity $T$. In (a) and (b), the solitonic motion in space ($x$) and time ($t/T$) is simulated via the numerical solutions of the dimensionless GP Eq.~(\ref{Eq:GPE2}). The impurity is numerically modeled as $-\gamma\delta(x)\approx -\frac{\gamma}{\sqrt{2\pi}\sigma}\exp(-x^2/(2\sigma^2))$, with $\gamma=0.1$ and $\sigma=0.002$. And the initial state $\psi(x,0)=Ae^{-(x-x_0)^2/l^2}e^{ikx}$, where $x_0=-3$, $l=0.4$, $k=0.1$, $\nu=0.1$ \color{black} and $A^2=15$. The color bar denotes the magnitude of $|\psi|^2$. The blue line in (a) denotes the center of the impurity potential. The green line in (b) denotes the pumping potential at $t=0$. Panels (c) and (d) present the results for the center-of-mass position of the soliton based on the Euler-Lagrangian Eqs.~(\ref{Scenter})-(\ref{Pbeta}) with scaled parameters $\tilde{\gamma}=\gamma/A=0.026$, and $\eta(0)=1$, $\kappa(0)=0.0258$, $\alpha(0)=\beta(0)=a(0)=0$. }\label{Fig:fig1}
     \end{figure}
In Eq.~(\ref{Eq:GPE1}), we consider the time-dependent superlattice potential $V_{\text{OSL}}/E_R=-V_{\text{s}}\cos^2\left(\frac{2\pi x}{d}\right)-V_\text{l}\cos^2\left(\frac{\pi x}{d}-\Omega t\right)$, with $E_R=\hbar^2\pi^2/(2m d^2)$. It consists of a primary lattice with the spatial period $d/2$ and lattice strength $V_{\text{s}}$, and a superimposed second lattice with the period $d$ and the strength $V_l$; the relative phase between the two lattices is $\Omega t$. Such a time-dependent potential has been experimentally realized in quantum gases to implement Thouless pumps~\cite{Lohse2016}. 

Moreover, we consider the impurity potential in Eq.~(\ref{Eq:GPE1}) as $V_{\text{imp}}/E_R=-V_0\delta(x)$, corresponding to an impurity centered at $x=0$ with the impurity strength $V_0$. Such an impurity trap has been experimentally realized using bichromatic fields~\cite{Fallani2007,Chomaz2023}, or `dark-state' optical potentials \cite{Lacki2016,Wang2018}.

For subsequent analysis, we shall rescale GP equation (\ref{Eq:GPE1}) into the dimensionless form via $x\rightarrow x/d$, $t\rightarrow 2E_Rt/\pi^2\hbar$, $\psi\rightarrow \psi\sqrt{|g|\pi^2/2E_R}$, $\gamma= V_0\pi^2/2d$, $V_{s(l)}=V_{s(l)}\pi^2/2$ and $\nu=\Omega\hbar\pi^2/2E_R$. The resulting dimensionless GP equation reads
    \begin{align}\label{Eq:GPE2}
	    i\frac{\partial}{\partial t}\psi=-\frac{1}{2}\frac{\partial^2}{\partial x^2}\psi-|\psi|^2\psi-\gamma\delta(x)\psi+V_{\text{OSL}}(x,t)\psi,
    \end{align}
  with the normalization condition $\int|\psi|^2dx=N$. The dimensionless pumping potential is explicitly given by
    \begin{align}
    	V_{\text{OSL}}(x,t)=-V_s\cos^2(2\pi x)-V_l\cos^2(\pi x-\nu t)\label{VOSL}
    \end{align}
  which is periodic in time with the periodicity $T=\pi/\nu$. In addition, the $\gamma$ in Eq.~(\ref{Eq:GPE2}) reflects the strength of the defect, and can be interpreted as the effective mass of the impurity~\cite{Jia2022}.

\section{Solitonic Thouless pumping through an impurity: numerical study} \label{NumericalGP}

In this section, we numerically study the soliton motion based on the GP equation~(\ref{Eq:GPE2}) under the initial excitation condition
$\psi(x,0)=Ae^{-(x-x_0)^2/l^2}e^{ikx}$ \cite{Qidong2022,Jia2022}, where $k$ and $x_0$ denote the initial momentum and the center-of-mass position, respectively; $l$ is the initial width of the envelope and $A$ is the peak amplitude. The ratio $\alpha=A^2/V_\textrm{min}$ with $V_\textrm{min}=\textrm{min}(V_s, V_l)$ effectively characterizes the strength of nonlinearity in the nonlinear Thouless pump. We will specialize to the regime of sufficiently strong nonlinearity, where stable soliton forms.

Before proceeding, we briefly review three familiar cases studied before. (i) In the absence of both impurity ($\gamma=0$) and Thouless pump ($V_{\text {OSL}}=0$), the quasi-1D BEC described by GP Eq.~(\ref{Eq:GPE2}) is known to support solitons as the nonlinear excitations. (ii) Without impurity ($\gamma=0$), Equation~(\ref{Eq:GPE2}) with $V_{\text {OSL}}\neq 0$ has been used to study the nonlinear Thouless pumping of solitons such as in Ref.~\cite{Qidong2022,Fu2022}. It is shown that for $\alpha$ below some critical value, i.e., in the pumped regime, the soliton undergoes topologically quantized transport, whereas for $\alpha$ above the critical value, i.e., in the trapped regime, the soliton is dynamically localized near its initial position. (iii) Equation~(\ref{Eq:GPE2}) with $\gamma\neq 0$ and $V_{\text {OSL}}=0$ has been used to study the interaction between a moving soliton with an impurity. As shown in Ref.~\cite{Forinash1994,Jia2022}, the behavior of the soliton crucially depends on the impurity mass $\gamma$. Specifically, the soliton can only move through a very light impurity, but becomes to be trapped when $\gamma$ becomes large. 

In this work, we are interested in the motion of the soliton when both the pump and the impurity are present, i.e., $V_{\text {OSL}}\neq 0, \gamma\neq 0$. To this end, henceforth we focus on the $\alpha$ in the pumped regime where solitons undergo topological transport in the defect-free case. We numerically solve the GP Eq.~(\ref{Eq:GPE2}) by the split-step fast Fourier algorithm~\cite{Muslu2004}; see Appendix A for technical details. In the numerical simulation, we approximate the $\delta$-potential by a sufficiently narrow Gaussian function according to the identity $
-\gamma\delta(x)=-\lim_{\sigma\rightarrow 0} \frac{\gamma}{\sqrt{2\pi}\sigma}e^{-x^2/(2\sigma^2)}$. For concreteness, we assume a pump with $V_s=V_l=25$, whose lowest band of the underlying band structure has the Chern number $C=1$. In addition, we choose the initial soliton width $l<0.5$ (i.e., smaller than the half spatial periodicity), so that the soliton state mainly occupies the lowest band when projecting onto the underlying linear Bloch states. Finally, the initial soliton momentum and the pump rate are chosen as $k=0.1$ and $\nu=0.1$, respectively, to ensure the adiabatic condition required by Thouless pump. Our numerical calculations reproduce the known results described earlier in (i)-(iii). The soliton motion under different parameters will be shown in Sec.~\ref{Discussion}.  

Our numerical results are shown in Figs.~\ref{Fig:fig1}(a) and (b) for a light impurity and in Figs.~\ref{Fig:fig2}(a) and (b) for a heavy impurity, respectively, where we compare the soliton behavior in the absence and presence of the topological pump.  Consider first the regime of light impurities [Fig.~\ref{Fig:fig1}(a)] without the pump. We see that the soliton directly tunnels through the light impurity potential centered at $x=0$, where the collision within the impurity only causes a small deviation of the soliton from its original path. In the presence of the topological pump with $V_s=V_l=25$, as shown in Fig.~\ref{Fig:fig1}(b), the soliton not only transmits through the impurity, but also the solitonic displacement becomes quantized to the underlying Chern number of the pump, despite the presence of the impurity.  

In comparison, Figs.~\ref{Fig:fig2}(a) and (b) illustrate the soliton motion for large impurity mass. Without the pump, as illustrated in Fig.~\ref{Fig:fig2}(a), we see that the soliton gets trapped in the impurity potential, as opposed to the case of light impurity in Fig.~\ref{Fig:fig1}(a). Interestingly, after applying Thouless pump with $V_s=V_l=25$ in Eq.~(\ref{VOSL}), the soliton is able to pass through the impurity instead of being localized by it [see Fig.~\ref{Fig:fig2}(b)]. Moreover, the solitonic displacement in a period becomes quantized to unity. This phenomena contrast strongly with the pump-free counterpart, and showcases the stability of the topological nonlinear pumping against the static disorder.

\section{Solitonic Thouless pumping through an impurity: variational approach} \label{VariationalGP}

To further establish our results and gain understandings how the interaction between the soliton and the impurity is affected by the Thouless pump, in this section we develop an analytical study of the soliton motion using the Lagrangian variational approach. 

We consider the renormalized wavefunction $\psi(x,t)=A\phi (X,\tau) e^{i(V_s+V_l)t/2}$ with the new variables $\tau=A^2t$ and $X=Ax$~\cite{Qidong2022,Fu2022}, and recast Eq.~(\ref{Eq:GPE2}) into the following form
\begin{equation}
		i\phi_{\tau}=-\frac{1}{2}\phi_{XX}-|\phi|^2\phi-\tilde{\gamma}\delta(X)\phi+\tilde{V}_{\text{OSL}}(X,\tau)\phi.\label{GPENP}
\end{equation}
Here, we have used the notations $\phi_\tau={\partial\phi}/{\partial \tau}$, $\phi_{XX}={\partial^2\phi}/{\partial X^2}$, and the scaled parameters $\tilde{V}_{\text{OSL}}(X,\tau)=-\tilde{V_s}\cos\left({4\pi X}/{A}\right)-\tilde{V_l}\cos\left({2\pi X}/{A}-{2\nu \tau}/{A^2}\right)$ with $\tilde{V}_s=V_s/(2A^2)$ and $\tilde{V_l}=V_l/(2A^2)$, and $\tilde{\gamma}=\gamma/A$. 

As the benchmark, we recall that without the pump ($\tilde{V}_{\text{OSL}}=0$), there exist two special cases where known solutions exist. 
(i) Without both the nonlinearity and the pump, Equation~(\ref{GPENP}) is reduced to the familiar
linear Schr\"odinger equation with the delta-potential, whose solution is a bound state localized in the defect potential (see e.g. Ref.~\cite{Forinash1994,Jia2022} and the references therein), i.e., $\phi_{\text{im}}(X)=\sqrt{\tilde{\gamma}}e^{-\tilde{\gamma}|X|}$ with the localization length $1/\tilde{\gamma}$. (ii) Without the impurity ($\tilde{\gamma}=0$) and the pump, Equation (\ref{GPENP}) supports moving solitons of the form
 $\phi_{\text{soliton}}=\eta\text{sech}(\eta(X-c\tau))\exp{(i(\eta^2-c^2)\tau/2+icX)}$~\cite{Forinash1994}, with the amplitude $\eta$, width $1/\eta$, and momentum $c$.

\begin{figure}[tb]
	\centering
	\includegraphics[width=1\columnwidth]{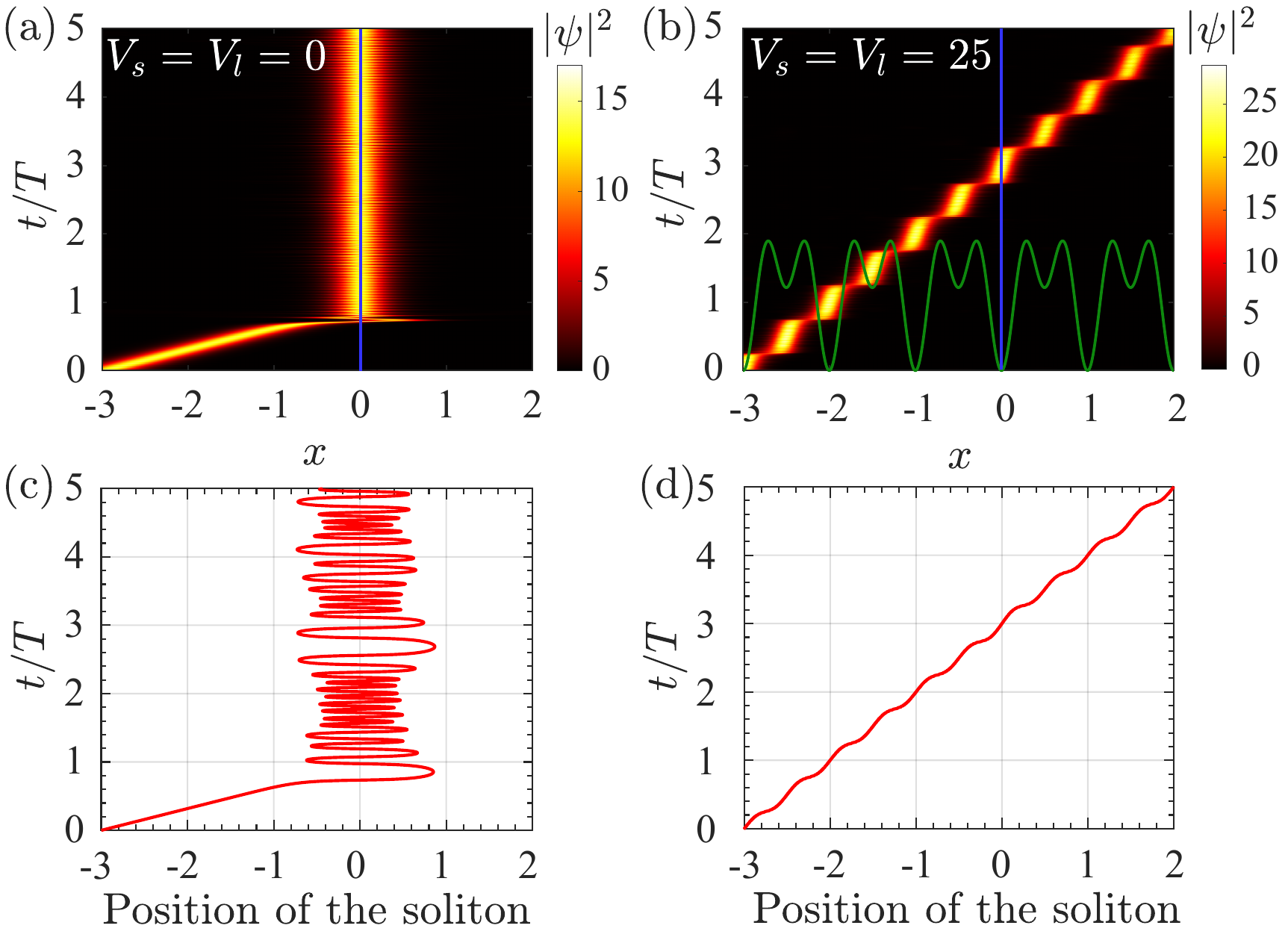}
	\caption{Soliton transport in the presence of a heavy impurity without and with nonlinear Thouless pumps. In (a) and (c), the pump is absent, where $V_s=V_l=0$. In (b) and (d), the pump is present, with $V_s=V_l=25$. In (a) and (b), the solitonic motion is simulated via the numerical solutions of the GP equation (\ref{Eq:GPE2}) where the impurity is approximated by $-\gamma\delta(x)\approx -\frac{\gamma}{\sqrt{2\pi}\sigma}\exp(-x^2/(2\sigma^2))$ with $\gamma=0.53$ and $\sigma=0.002$. The initial condition is $\psi(x,0)=Ae^{-(x-x_0)^2/l^2}e^{ikx}$, where $x_0=-3$, $l=0.4$, $k=0.1$, $\nu=0.1$ and $A^2=15$. Panels (c) and (d) present the results for the center-of-mass position of the soliton based on the Euler-Lagrangian equation Eqs. (\ref{Scenter})-(\ref{Pbeta}) with scaled parameters $\tilde{\gamma}=0.137$ and $\eta(0)=1$, $\kappa(0)=0.0258$, $\alpha(0)=\beta(0)=a(0)=0$.}\label{Fig:fig2}
\end{figure}
In the presence of the Thouless pump ($\tilde{V}_{\text{OSL}}(X,\tau)\neq 0$), neither the bound state $\phi_{\text{im}}$ nor the soliton $\phi_{\text{soliton}}$ stands as the general solutions of Eq.~(\ref{GPENP}). To seek the solutions in this case, we use the Lagrangian variational approach along the line of Refs.~\cite{Forinash1994,Jia2022}. We assume a trial wave function which is the combination of the moving soliton and the impurity-induced bound state, i.e., 
    \begin{equation}
\!\!\! \!\!	\phi(X,\tau)=\left[\eta\text{sech}(\eta X-z)e^{i\kappa X}+a\sqrt{\tilde{\gamma}}e^{-\tilde{\gamma}|X|+i\alpha}\right]e^{i\beta},\label{ansatz}
    \end{equation}
\noindent where $\eta(\tau)$, $z(\tau)$, $\kappa(\tau)$, $a(\tau)$, $\alpha(\tau)$ and $\beta(\tau)$ are the variational parameters to be determined below. Specifically, $\eta(\tau)$ and $z(\tau)$ are the amplitude and the center-of-mass position of the soliton, respectively, $\kappa(\tau)$ is the wavenumber of the soliton, $a(\tau)$ is the amplitude of the bound-state component, $\alpha(\tau)$ is the relative phase between the soliton and the bound state, and $\beta(\tau)$ is the global phase of the trial function. The key assumption underlying the ansatz (\ref{ansatz}) is that the functional forms of the soliton and the impurity-induced bound state are preserved in the presence of the Thouless pumping, but the corresponding parameters become slowly time-dependent. 

With Eq.~(\ref{ansatz}), the Lagrangian of $L$ corresponding to the GP Eq.~(\ref{GPENP}) can be derived following standard procedures~\cite{Jia2022,Forinash1994}. The Lagrangian is written as
\begin{equation}
		L=\frac{i}{2}\int_{-\infty}^{+\infty}\left(\phi^\star\frac{\partial\phi}{\partial \tau}-\phi\frac{\partial\phi^\star}{\partial \tau}\right)dX-E, \label{Ld}
	\end{equation}
	where we have
	\begin{eqnarray}
		E&=&\frac{1}{2}\int_{-\infty}^{+\infty}\left|\frac{\partial \phi}{\partial x}\right|^2dX-\frac{1}{2}\int_{-\infty}^{+\infty}|\phi|^4dX\notag\\
		&-&\int_{-\infty}^{+\infty}\tilde{\gamma}\delta(X)|\phi|^2dX-\int_{-\infty}^{+\infty}\tilde{V}_s\cos\left(\frac{4\pi X}{A}\right)|\phi|^2dX\notag\\
		&-&\int_{-\infty}^{+\infty}\tilde{V}_l\cos\left(\frac{2\pi X}{A}-\frac{2\nu \tau}{A^2}\right)|\phi|^2dX. \nonumber
	\end{eqnarray}
Inserting ansatz (\ref{ansatz}) into Eq.~(\ref{Ld}), and after some tedious calculation, we obtain $L=L_1+L_2+L_3+L_4+L_5$ as the summation of following terms:
		\begin{eqnarray}
		L_1&=&\frac{i}{2}\int_{-\infty}^{+\infty}\left(\phi^\star\frac{\partial\phi}{\partial \tau}-\phi\frac{\partial\phi^\star}{\partial \tau}\right)dX\notag\\
		&=&-(2\eta+a^2)\frac{\partial\beta}{\partial \tau}-2z\frac{\partial\kappa}{\partial\tau}-a^2\frac{\partial \alpha}{\partial \tau}.\nonumber
	\end{eqnarray}
	The term associated with kinetic energy is calculated as 
	\begin{eqnarray}
		L_2&=&-\frac{1}{2}\int_{-\infty}^{+\infty}\left|\frac{\partial\phi}{\partial x}\right|^2dX\nonumber\\
		&=&-\left(\kappa^2\eta+\frac{1}{3}\eta^3+\frac{1}{2}a^2\tilde{\gamma}^2\right).\nonumber
	\end{eqnarray}
	The nonlinear term is evaluated as
	\begin{eqnarray}
L_3&=&\frac{1}{2}\int_{-\infty}^{+\infty}|\phi|^4dX\nonumber\\
		&=&\frac{2}{3}\eta^3+\frac{1}{4}a^4\tilde{\gamma}. \nonumber
	\end{eqnarray}
	The contribution from the impurity is given by
	\begin{eqnarray}
		\!\!\!\!\!\!\!\!\!\!		L_4&=&\int_{-\infty}^{+\infty}\tilde{\gamma}\delta(X)|\phi|^2dX\notag\\
		&=&\tilde{\gamma}\left[\eta^2\text{sech}^2(z)+2a\eta\text{sech}(z)\tilde{\gamma}^{1/2}\cos(\alpha)+a^2\tilde{\gamma}\right]. \nonumber
	\end{eqnarray}
	The term associated with the pump reads as
	\begin{eqnarray}
		L_5&=&\int_{-\infty}^{+\infty}\left[\tilde{V}_s\cos\left(\frac{4\pi X}{A}\right)+\tilde{V}_l\cos\left(\frac{2\pi X}{A}-\frac{2\nu \tau}{A^2}\right)\right]|\phi|^2dX\notag\\
		&=&\tilde{V}_s\left[\frac{4\pi^2}{A}\cos\left(\frac{4\pi z}{A\eta}\right)\text{csch}\left(\frac{2\pi^2}{A\eta}\right)+\frac{a^2\tilde{\gamma}^2A^2}{4\pi^2+A^2\tilde{\gamma}^2}\right]\nonumber\\
		&+&\frac{2\pi^2\tilde{V}_l}{A}\cos\left(\frac{2\pi A z-2\eta\nu\tau}{A^2\eta}\right)\text{csch}\left(\frac{\pi^2}{A\eta}\right)\notag\\
		&+&\frac{a^2\tilde{\gamma}^2A^2\tilde{V}_l}{\pi^2+A^2\tilde{\gamma}^2}\cos\left(\frac{2\nu \tau}{A^2}\right). \nonumber
	\end{eqnarray}
In the above derivations, we have neglected all the terms associated with the overlapping of the two components (i.e., solitonic component and the  bounded component) of the ansatz. Physically, this means that we ignore the interaction between the soliton and the impurity when they are far from each other, while retaining their interaction when the soliton is close to the impurity~\cite{Forinash1994}.

Finally, after some rearrangements, the Lagrangian is written as~
\begin{widetext}
 \begin{eqnarray}
\!\!\!\!\!\!\!\!\!\!     	L&=&\!-\!2\eta\frac{\partial\beta}{d\tau}\!-\!2z\frac{\partial\kappa}{\partial\tau}-\!\kappa^2\eta\!+\!\frac{1}{3}\eta^3\!+\!{2\pi^2}\left[\frac{2\tilde{V_s}}{A}\cos\left(\frac{4\pi z}{\eta A}\right)\text{csch}\left(\frac{2\pi^2}{\eta A}\right)+\frac{\tilde{V}_l}{A}\cos\left(\frac{2\pi A z-2\eta\nu \tau}{\eta A^2}\right)\text{csch}\left(\frac{\pi^2}{\eta A}\right)\right]\nonumber\\
	&+&\!\!\tilde{\gamma}\eta^2\text{sech}^2(z)\!+2(a\tilde{\gamma}^{3/2})\eta \text{sech}(z)\cos(\alpha)-\!a^2\left(\frac{\partial\beta}{d\tau}\!+\frac{\partial\alpha}{\partial\tau}\right)+\frac{a^2\tilde{\gamma}^2}{2}+\frac{a^2\tilde{\gamma}^2A^2\tilde{V}_s}{4\pi^2+A^2\tilde{\gamma}^2}+\frac{a^2\tilde{\gamma}^2A^2\tilde{V}_l}{\pi^2+A^2\tilde{\gamma}^2}\cos\left(\frac{2\nu \tau}{A^2}\right)+\frac{1}{4}a^4\tilde{\gamma}.\label{Lagrang}
\end{eqnarray}
The first line of Lagrangian~(\ref{Lagrang}) does not depend on impurity parameters, corresponding to the contribution from the pumped soliton in the absence of impurity. The second line involve the corrections due to impurities up to the fourth orders of $a$ associated with the weight of bound-state component in Eq.~(\ref{ansatz})~\cite{Forinash1994}. In particular, the terms in order of $O(a^1)$ come from the mutual interaction between the bound state and the soliton; the terms in order of $O(a^2)$ describe the impurity's energy in the presence of the pumping potential; the terms in order of $O(a^4)$ arise from the self-interaction of the bound state.

Thus the Euler-Lagrangian equations for the variational parameters are given by $\partial L/\partial q_i-d/dt(\partial L/\partial \dot{q_i})=0$ with $q_i=\eta, z, \kappa, a, \alpha, \beta$. After some tedious but straightforward calculations, we obtain Eqs.~(\ref{Scenter})-(\ref{Pbeta}) below. Specifically, crucial for our study is the center-of-mass position $z(\tau)$ of the soliton, whose equation of motion are given by	
	\begin{eqnarray}
\frac{d^2z}{d\tau^2}=&-&a\eta\kappa\tilde{\gamma}^{3/2}\text{sech}(z)\sin(\alpha)-\frac{8\pi^3\tilde{V_s}}{A^2}\text{csch}\left(\frac{2\pi^2}{A\eta}\right)\sin\left(\frac{4\pi z}{A\eta}\right)
-\frac{2\pi^3\tilde{V_l}}{A^2}\text{csch}\left(\frac{\pi^2}{A\eta}\right)\sin\left(\frac{2A\pi z-2\nu \tau\eta}{A^2 \eta}\right)\nonumber\\
&-&\tilde{\gamma}\eta^3\text{sech}^2(z)\text{tanh}(z)-a\eta^2\tilde{\gamma}^{3/2}\cos(\alpha)\text{sech}(z)\text{tanh}(z).\label{Scenter}
    \end{eqnarray}	
The dynamics of $z(\tau)$ depends on the time-dependent parameters $\kappa$, $\eta$, $a$, $\alpha$ and $\beta$, whose equations of motion are given as follows
    \begin{eqnarray}
    	\frac{d}{d\tau}a&=&\eta\tilde{\gamma}^{3/2}\text{sech}(z)\sin(\alpha),\label{Eq:a}\\
	\frac{d}{d\tau}\eta&=&-a\eta\tilde{\gamma}^{3/2}\text{sech}(z)\sin(\alpha), \label{Pa}\\
    \frac{d}{d\tau}\kappa&=&-\frac{8\pi^3\tilde{V_s}}{A^2\eta}\text{csch}\left(\frac{2\pi^2}{A\eta}\right)\sin\left(\frac{4\pi z}{A\eta}\right)-\frac{2\pi^3\tilde{V_l}}{A^2\eta}\text{csch}\left(\frac{\pi^2}{A\eta}\right)\sin\left(\frac{2A\pi z-2\nu \tau\eta}{A^2 \eta}\right)-\tilde{\gamma}\eta^2\text{sech}^2(z)\text{tanh}(z)\nonumber\\
    &-&a\eta{\tilde{\gamma}}^{3/2}\cos(\alpha)\text{sech}(z)\text{tanh}(z),\label{Pkappa}\\
     \frac{d}{d\tau}\alpha&=&\frac{1}{a}\cos(\alpha)\text{sech}(z)\eta{\tilde{\gamma}}^{3/2}+\frac{1}{2}a^2\tilde{\gamma}-\frac{d}{d\tau}\beta+\frac{\tilde{\gamma}^2}{2}+\frac{A^2\tilde{V}_l\tilde{\gamma}^2}{\pi^2+A^2\tilde{\gamma}^2}\cos\left(\frac{2\nu \tau}{A^2}\right)+\frac{A^2\tilde{V}_s\tilde{\gamma}^2}{4\pi^2+A^2\tilde{\gamma}^2},\label{Eq:alpha}\\
     	\frac{d}{d\tau}\beta&=&\frac{4\pi^4\tilde{V}_s}{A^2\eta^2}\cos\left(\frac{4\pi z}{A\eta}\right)\text{coth}\left(\frac{2\pi^2}{A\eta}\right)\text{csch}\left(\frac{2\pi^2}{A\eta}\right)+\frac{8\pi^3\tilde{V}_s}{A^2\eta^2}\text{csch}\left(\frac{2\pi^2}{A\eta}\right)\sin\left(\frac{4\pi z}{A\eta}\right)z\nonumber\\
     	&+&\frac{\pi^4\tilde{V_l}}{A^2\eta^2}\cos\left(\frac{2A\pi z-2\tau\nu\eta}{A^2\eta}\right)\text{coth}\left(\frac{\pi^2}{A\eta}\right)\text{csch}\left(\frac{\pi^2}{A\eta}\right)+\frac{2\pi^3\tilde{V}_l}{A^2\eta^2}\text{csch}\left(\frac{\pi^2}{A\eta}\right)\sin\left(\frac{2A\pi z-2\tau\nu\eta}{A^2\eta}\right)z\nonumber\\     	
     	&+&\tilde{\gamma}\text{sech}^2(z)\eta+\frac{1}{2}\eta^2-\frac{\kappa^2}{2}+a{\tilde{\gamma}}^{3/2}\cos(\alpha)\text{sech}(z).\label{Pbeta}
    \end{eqnarray}
\end{widetext}

Equation~(\ref{Scenter}) provides the key analytical result of this work; it shows how the motion of a soliton against an impurity is affected by the Thouless pump. Note that without the pump (i.e., $\tilde{V}_s=\tilde{V}_l=0$), the above equations recover the corresponding results of Ref.~\cite{Forinash1994}. As shown in Eq.~(\ref{Scenter}), the displacement of the soliton is strongly affected by the time-dependent superposition amplitudes of the soliton component $\eta(\tau)$ and the bound-state component $a(\tau)$ in Eq.~(\ref{Pa}). 

\begin{figure}[tb]
	\centering
	\includegraphics[width=1\columnwidth]{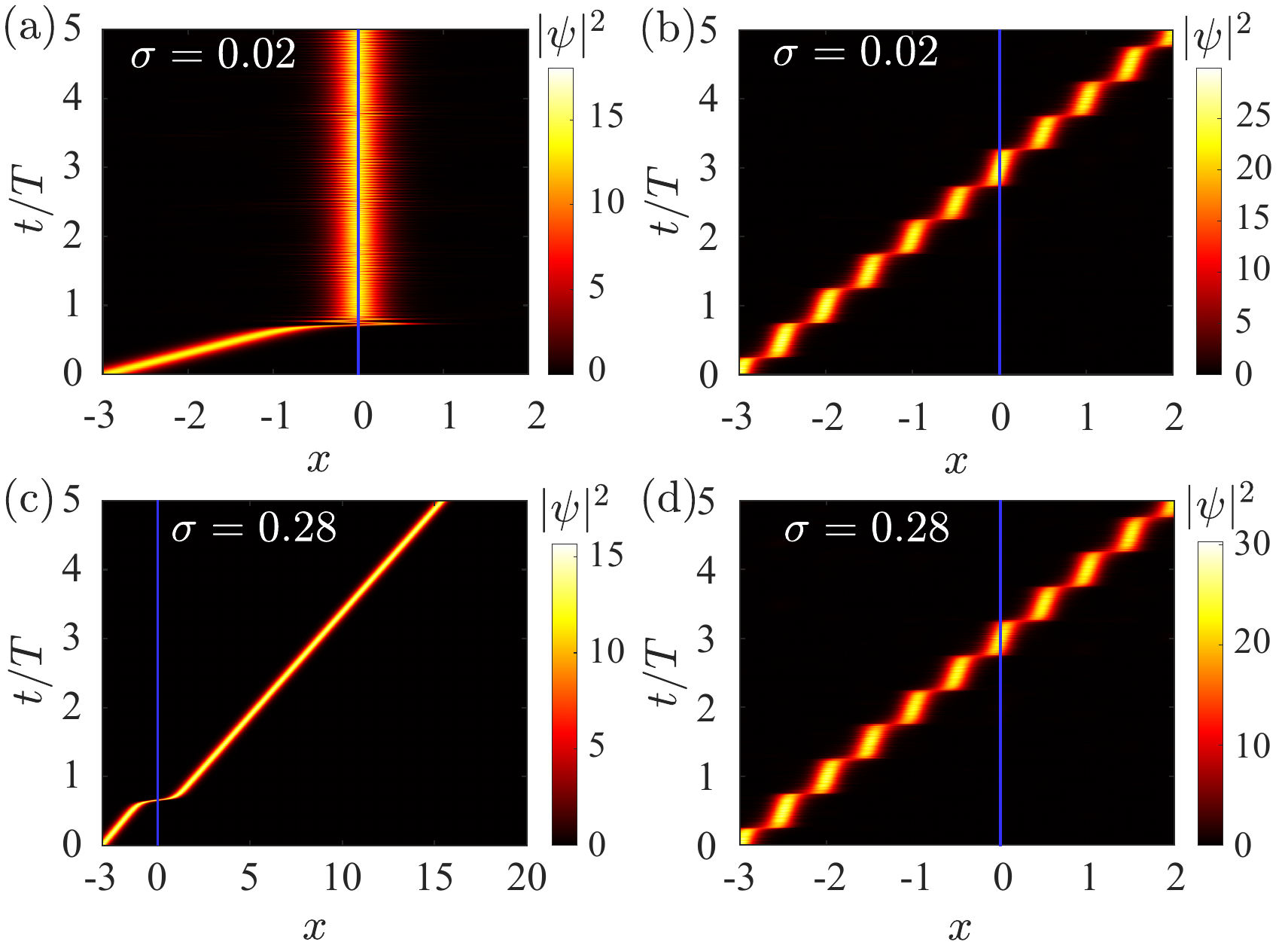}
	\caption{Numerical simulations of soliton transport in the presence of a Gaussian impurity $-\frac{\gamma}{\sqrt{2\pi}\sigma}e^{-x^2/(2\sigma^2)}$ with $\gamma=0.53$, for the finite width $\sigma=0.02$ in panels (a) and (b), and $\sigma=0.28$ in (c) and (d). Numerical computations are performed in free space ($V_s=V_l=0$) in (a) and (c), and with Thouless pump ($V_s=V_l=25$) in (b) and (d). Other parameters are the same as Fig.~\ref{Fig:fig2}(b).}\label{Fig:fig3}
\end{figure}

\begin{figure}[tb]
	\centering
	\includegraphics[width=1\columnwidth]{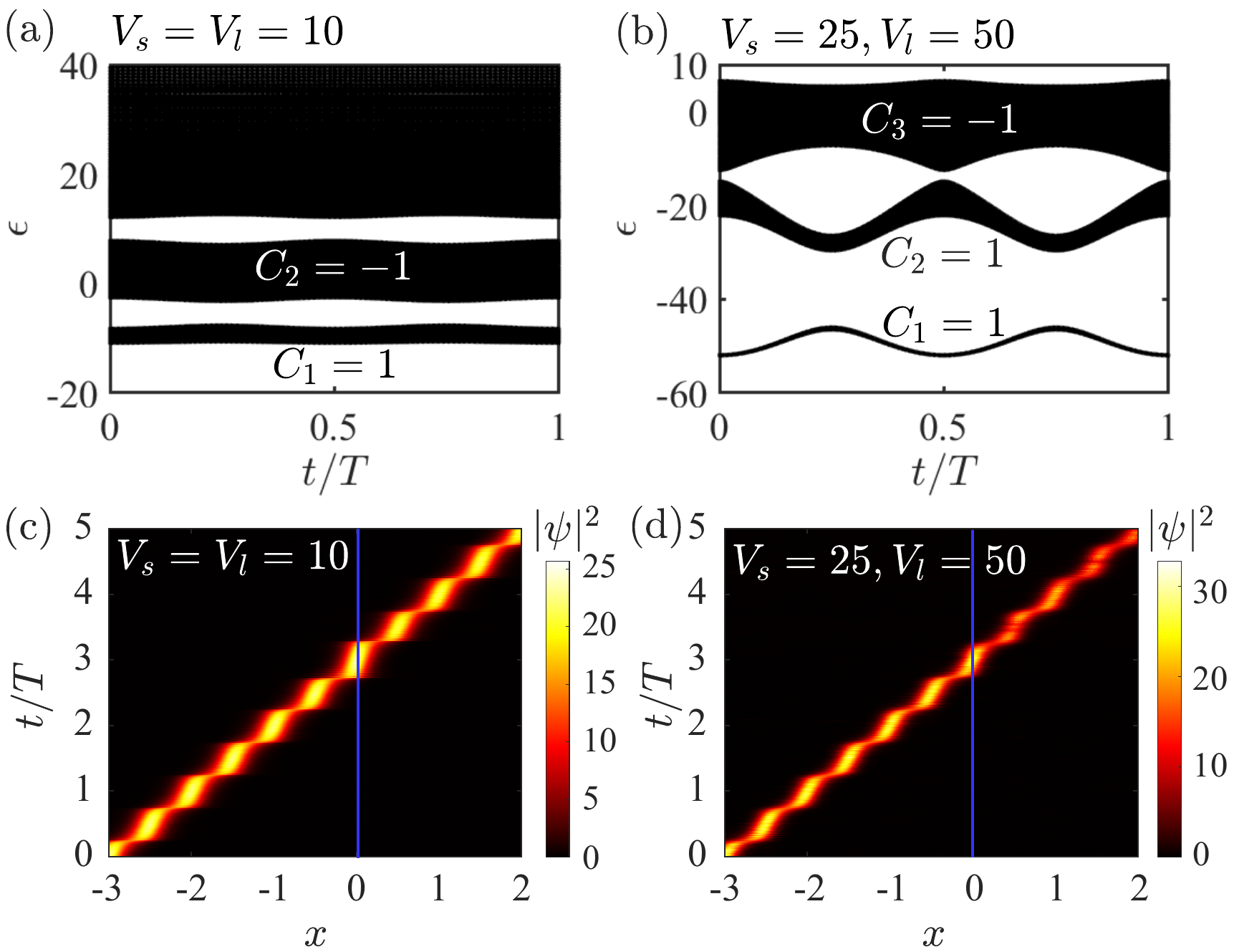}
	\caption{Numerical simulations of soliton transport in the presence of a heavy impurity, for Thouless pump with $V_s=V_l=10$ in (a) and (c), and $V_s=25$ and $V_l=50$ in (b) and (d). Panels (a) and (b) show the corresponding band structures associated with the linear pump. The lowest two bands have Chern numbers (a) $C=\{1,-1\}$, and (b) $C=\{1,1\}$. Panels (c) and (d) show the corresponding soliton motion. Other parameters are same as Fig.~\ref{Fig:fig2}(b).}\label{Fig:fig4}
\end{figure}

To solve the set of Euler-Lagrangian equations in Eqs.~(\ref{Scenter})-(\ref{Pbeta}) for the variational parameters, without loss of generality, we consider the following initial conditions. At $\tau=0$, the position of the bright soliton is centered at $z(0)/A=-3$, far away from the impurity; the initial amplitude and velocity of the soliton are chosen as $\eta=1$ and $\kappa= 0.0258$, respectively; for other parameters, we choose $a(0)=\alpha(0)=\beta(0)=0$. 

To double-check whether our derivations reproduce the known results in Ref.~\cite{Forinash1994} in the absence of Thouless pumps, at first, we take $\tilde{V}_s=0$ and $\tilde{V}_l=0$ in Eqs.~(\ref{Scenter})-(\ref{Pbeta}). Indeed, as illustrated in Fig.~\ref{Fig:fig1}(c) for the light impurity with $\tilde{\gamma}=0.026$, the soliton directly transits through the impurity potential without being localized. This can be intuitively understood by noting that for small $\tilde{\gamma}$, the bound state has a much smaller energy than the soliton, resulting in a small admixture of the localized bound-state in Eq.~(\ref{ansatz}). Therefore, the moving soliton is only slightly perturbed by the impurity. With the increase of $\tilde{\gamma}$, however, the bound state gains more weight in the wavefunction described by Eq.~(\ref{ansatz}). For $\tilde{\gamma}=0.137$ as in Fig.~\ref{Fig:fig2}(c), we see that the initially moving soliton becomes localized around the impurity, consistent with previous findings. 

Next, we turn on the Thouless pump with $\tilde{V}_s=\tilde{V}_l=0.83$ (corresponding to ${V}_s={V}_l=25$ and $A^2=15$), and again solve Eqs.~(\ref{Scenter})-(\ref{Pbeta}). The result for the light impurity is shown in Fig.~\ref{Fig:fig1}(d). We see that the soliton in the Thouless pump pass through a light impurity as if the impurity does not exist, and its mean displacement in a period becomes quantized. Remarkably, as shown in Fig.~\ref{Fig:fig2}(d), the nonlinear topological pump enables the soliton to pass through even a very heavy impurity, contrary to the pump-free case, and the motion is quantized. This provides a direct evidence that the quantized transport of a quantum particle in nonlinear topological Thouless pump is robust against the local imperfections.

\section{Discussion} \label{Discussion}
In previous sections, we have illustrated the robustness of the nonlinear Thouless pumping of solitons against an $\delta$-impurity when $A^2=15$ and $V_s=V_l=25$. In this section, we extend our discussions to general conditions of the impurity, the soliton and the lattice, respectively. 

As mentioned earlier, numerically, the $\delta$-impurity is approximated by a narrow Gaussian function $-\gamma\delta(x)=-\lim_{\sigma\rightarrow 0}\frac{\gamma}{\sqrt{2\pi}\sigma}e^{-x^2/(2\sigma^2)}$. For instance, in Fig.~\ref{Fig:fig2}(b) with $\gamma=0.53$, a sufficiently small impurity width $\sigma=0.002$ was chosen, so that without both the pump and nonlinearity the numerical solution of the system's ground state agrees perfectly with the exact solution of the bound state $\sqrt{\gamma}e^{-\gamma |x|}$. In this limit, we see that the soliton, arrested by the impurity in free space, can undergo a quantized transport through the impurity. Increasing $\sigma$ away from the $\delta$-function limit, as in a more realistic defect with finite width, the soliton in free space remains trappable by the impurity with the width $\sigma\ll l$ smaller than the soliton width $l$, as shown in Fig.~\ref{Fig:fig3}(a) for $\sigma=0.02$ and $l=0.4$. However, for $\sigma\gtrsim l$, the soliton can transmit through the impurity even without the pump [Fig.~\ref{Fig:fig3}(c)]. In both cases, adding Thouless pump results in topologically quantized soliton motion through the impurity [Figs.~\ref{Fig:fig3}(b) and (d)]; the average displacement is quantized to the Chern number of the lowest band.

Changing the lattice strength $V_s$ and $V_l$ results in modifications of the underlying band structures. Figure~\ref{Fig:fig4}(a) considers a decreased lattice strength, $V_s=V_l=10$, while other parameters are the same as in Fig.~\ref{Fig:fig2}(b). Although the energy gap becomes smaller, the topology of the relevant bands remains the same as for $V_s=V_l=25$. Thus, the soliton is expected to display similar behavior as in Fig.~\ref{Fig:fig2}(b), as seen in Fig.~\ref{Fig:fig4}(c). Figure~\ref{Fig:fig4}(b) shows the band structure associated with the linear pump with $V_s=25$ and $V_l=50$, where the Chern number of the lowest band remains to be $C=1$, but the Chern number of the second band is different from Fig.~\ref{Fig:fig4}(a). Since the given soliton mainly occupies the lowest band, still, similar topological soliton transmission across the impurity is observed [Fig.~\ref{Fig:fig4}(d)], where the average displacement is predicted by the Chern number $C=1$ of the lowest band. 

\begin{figure}[t!]
	\centering
	\includegraphics[width=1\columnwidth]{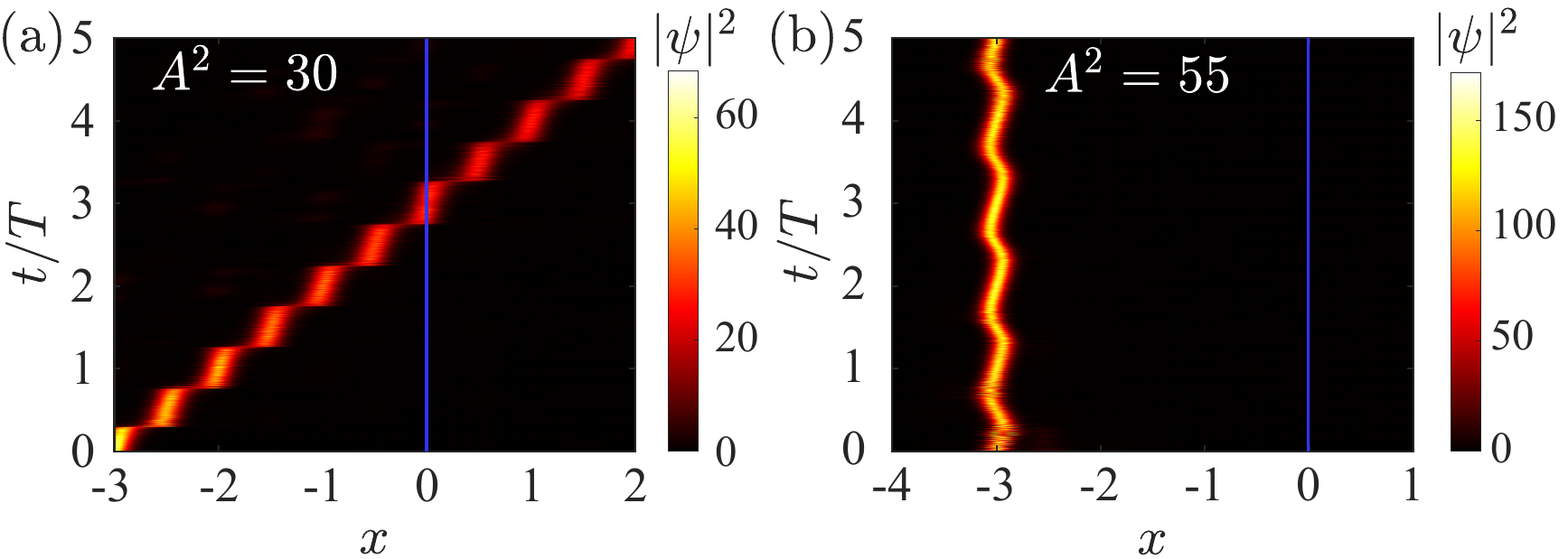}
	\caption{Numerical simulations of soliton transport in the presence of a heavy impurity for (a) $A^2=30$ and (b) $A^2=55$. Other parameters are the same as in Fig.~\ref{Fig:fig2}(b).}\label{Fig:fig5}
\end{figure}

The topological transmission through the impurity is generally observed when the strength of the nonlinearity is in the pumped regime [Fig.~\ref{Fig:fig5}(a)]. For sufficiently strong nonlinearity above the critical value, the soliton becomes to be trapped near its initial position away from the impurity, as shown in Fig.~\ref{Fig:fig5}(b).

Over all, our above results suggest that, as same as the linear case, the nonlinear topological pump is immune to microscopic imperfections.

	\section{Conclusion}\label{Con}
 In this work, we theoretically investigate the nonlinear Thouless pumping of a soliton in the presence of an impurity based on the system of a quasi-1D BEC. Using both the GP equation and the Lagrangian variational method, we compare the interaction between the impurity and the soliton in the absence and presence of the nonlinear topological pumps. Without the pump, whether the soliton can move through the impurity depends crucially on the impurity mass. In contrast, the soliton in the nonlinear Thouless pump can transmit through even a heavy impurity and exhibits a quantized motion. This work sheds light on the interplay of nonlinearity, topology and disorder, and promises potential applications in information processing with solitons.
	
	We thank Qidong Fu, Fangwei Ye, Qi Zhang, and Biao Wu for stimulating discussions. This work was supported by the National Natural Science Foundation of China (Nos. 12074344,12374246), the Zhejiang Provincial Natural Science Foundation (Grant Nos. LZ21A040001) and the key projects of the Natural Science Foundation of China (Grant No. 11835011).

	\appendix
		
	\section{Numerical method}
	
	We numerically solve the dimensionless GP Eq.~(\ref{Eq:GPE2}) of the main text by the split-step fast Fourier algorithm (see e.g., Ref.~\cite{Muslu2004} and the references therein). In this section, we present details on our numerical techniques. 
	
	By approximating $-\gamma\delta(x)=-\frac{\gamma}{\sqrt{2\pi}\sigma}e^{-x^2/(2\sigma^2)}$, we rewrite Eq.~(\ref{Eq:GPE2}) of the main text as 
	\begin{align}\label{Eq:f1f2GP}
		i\frac{\partial \psi}{\partial t}=(O_1+O_2)\psi,
	\end{align}
	where the two operators are 
	\begin{eqnarray}
		O_1&=&-\frac{1}{2}\frac{\partial^2}{\partial x^2},\nonumber\\
		O_2&=&V_{\text{OSL}}(x,t)-|\psi|^2-\frac{\gamma}{\sqrt{2\pi}\sigma}e^{-x^2/(2\sigma^2)}. \nonumber
	\end{eqnarray}
We have $[O_1, O_2]\neq 0$. To solve the time evolution, we implement the first-order splitting scheme 
	\begin{equation}
		\psi(x,t+dt)=\exp(-iO_1dt)\exp(-iO_2dt)\psi(x,t)\label{Eq:FirstAppro}
	\end{equation}
	with a sufficiently small time step $dt=10^{-4}$. We use the fast Fourier transform (FFT) algorithm to compute Eq.~(\ref{Eq:FirstAppro}): Let $\mathcal{F}$ denotes the Fourier transform and $\mathcal{F}^{-1}$ denotes the inverse Fourier transform, we have
	\begin{align}\label{Eq:SSFMFinalEq}
		\psi(x,t+dt)=\mathcal{F}^{-1}\left[\exp\left(-\frac{i}{2}k^2dt\right)\mathcal{F}\left[\exp(-i O_2dt)\psi(x,t)\right]\right].
	\end{align}
Finally, we choose $dx=5\times 10^{-4}$ in the spatial discretization of the GP equation. 

	Using the above numerical scheme, we have reproduced the known results of soliton motion for $\gamma=0$ in the absence of impurity, as well as the exact solutions of the bound state when both the nonlinearity and the pump are absent. In the presence of both the pump and the impurity, our numerical simulations of the soliton motion find good agreement with the results obtained from the variational approach.

\bibliography{Reference}

\end{document}